# Super-sensitive Molecule-hugging Graphene Nanopores

Slaven Garaj[1], Song Liu[1,2], Daniel Branton[3], and Jene A. Golovchenko[1,4]*

[1]Department of Physics, Harvard University, Cambridge Massachusetts, 02138, USA

[2]Department of Physics, Peking University, Beijing 100871, P.R. China

[3]Department of Molecular and Cellular Biology, Harvard University, Cambridge, Massachusetts, 02138, USA

[4]School of Engineering and Applied Sciences, Harvard University, Cambridge, Massachusetts, 02138, USA

*Corresponding author, email: golovchenko@physics.harvard.edu

**Longitudinal resolution and lateral sensitivity are decisive characteristics that determine the suitability of a nanopore sensor for sequencing a strand of DNA as well as other important polymers. Previous modeling of DNA induced ionic current blockades in single atom thick graphene nanopores has shown these nanopores to have sufficient longitudinal resolution to distinguish individual nucleobases along the length of a DNA molecule[1]. Here we experimentally focus on the sensitivity to small changes in DNA diameter that can be discerned with graphene nanopores. We show that remarkably large sensitivities (*0.5 nA/Å*) are obtained when the nanopore is tailored to have a diameter close to that of the polymer of interest. Our results have been obtained with double-stranded DNA (dsDNA). Smaller graphene nanopores that can be tuned to the diameter of single-stranded DNA (ssDNA) for sequencing have only recently been demonstrated[2]. Our results indicate that nanopore sensors based on such pores will provide excellent resolution and base discrimination, and will be robust competitors to protein nanopores that have recently demonstrated sequencing capability[3,4].**

Biological[5] and solid-state[6] nanopore sensors are becoming key elements in next generation DNA sequencing technologies[7,8]. They are increasingly being used for rapid characterization of many individual DNA molecules[9-11], proteins[12,13], and protein-DNA complexes[14,15]. In a nanopore sensor, a charged biopolymer in solution is electrophoretically driven through a nanometer scale pore in an insulating membrane separating two voltage biased reservoirs filled with aqueous ionic solution. As the negatively charged DNA molecules pass through a nanopore, they block the flow of ions through the nanopore, leading to a transient drop of the ionic current through the nanopore (a blockade). The instantaneous ionic current during the blockade is related to the geometrical and chemical properties of the pore and the part of the polymer in the pore at that given time. Here we demonstrate an extraordinary sensitivity of graphene nanopores to DNA molecules when the pore and molecule diameters are very closely matched.

We fabricated single nanopores in graphene membranes[1], see also[16,17], with a focused beam of a 200 KeV transmission electron microscope (TEM). The graphene membranes were suspended over 200 nm x 200 nm apertures in thin, free-standing $SiN_x$ films supported on silicon frames (Fig. 1a and Methods section). DNA translocation through these graphene nanopores was investigated in high salt and high pH solutions (3M KCl, pH 10). Under these conditions electrostatic and chemical stick-slip forces between DNA molecules and the graphene surface are largely suppressed as will be demonstrated below. The applied voltage bias between the reservoirs on either side of the graphene membrane was 160 mV.

Figure 1b shows individual current blockades in graphene nanopores of various diameters induced by 10kb dsDNA molecules. A strong nanopore diameter dependence is evident, with blockade currents that are much larger than have been observed in $SiN_x$ nanopores (Fig. 1b and c) or protein nanopores. This is especially clear for the nanopores whose diameters approach the size of the translocating polymer. As we show below, such polymer-hugging nanopores are extraordinarily sensitive to small changes in the difference between the molecule and nanopore diameter.

Figure 1d shows ionic conductivity vs. nanopore diameter for a series of open graphene nanopores. The linear dependence of conductivity on diameter is expected only when the membrane is much thinner than the pore diameter[1]. This is consistent with the predicted access resistance of a small circular pore[18], our numerical solutions of the Laplace equation for the ionic current density[1], and molecular dynamic calculations[19].

To understand the sensitivity of graphene nanopores to single molecule DNA translocation, we investigated many DNA translocation events for pores of different diameters (Figs. 2a & b). In nanopores with diameters greater than ~4 nm, dsDNA molecules translocate through the pore either as extended linear molecules (leading to a single-dip event, Fig. 2c, left trace) or as a folded molecule (typically doubling the current blockade during the folded part of the blockade (Fig. 2c, right trace). In nanopores with diameters ≤3.2 nm, only unfolded molecules were observed (single-dip events), consistent with the notion that these nanopores are too small to admit folded dsDNA molecules for passage. We characterize each DNA translocation event by two parameters: the event duration, which indicates the time it takes for the molecule to fully translocate through the nanopore, and the average current blockade magnitude, which reflects the extent to which the molecule obstructs or blocks the flow of ionic current through the nanopore. The product of these two parameters is the electronic charge deficit[10] (*ecd*) during the event. Note the clustering of all the dsDNA translocation events around the hyperbolic line of constant *ecd* in Fig. 2a. This, together with the short average translocation time ($T_m$~150μs), shows that dsDNA molecules in 3M KCl, pH 10, freely translocate through the graphene nanopore and are minimally retarded by stochastic sticking interactions with the graphene membrane or nanopore[10].

Raising the pH of the salt solution from 10 to 12.5 denatures the dsDNA into single-stranded DNA (ssDNA) molecules[11]. Unfolded ssDNA translocation events (Fig. 2d right) show blockade magnitudes less than half that of dsDNA (Fig. 2d, left trace), a reflection of the reduced diameter of ssDNA. To our knowledge this is the first observation of ssDNA translocating through a graphene nanopore. Many ssDNA events are observed far from the line of constant *ecd* because of their extended translocation times (Fig. 2c). This strongly suggests that many of the ssDNA molecules stick to, or interact, with graphene. This is not surprising since the nucleobases in ssDNA are accessible and known to form π-π stacking/van der Waals interactions with a graphene surface[20].

The average current blockade for unfolded dsDNA translocations in graphene nanopores was measured as a function of nanopore diameter, and the results are shown as the filled circles in Fig. 3a. A steep increase in blockade current with decreasing nanopore diameter, from 6.5 to 2.7 nm, is observed. This indicates a continuing increase in sensitivity to the difference in molecule diameter and pore diameter as the fit between the two gets closer and closer. The open circles are data obtained from e-beam drilled nanopores in 30 nm thick $SiN_x$ which, because of the membrane thickness, show virtually no change in blockade current with nanopore diameter.

To understand the experimental data in Fig. 3a, we calculated (see Methods) current flow through an open pore ($I_0$) vs. pore diameter, and through the same pore containing a DNA molecule passing through its center, to obtain the blockade current ($I_B$). The electrolyte solution is modeled as a continuous medium with appropriate ionic conductivity, and we numerically solved the Laplace equation with appropriate boundary conditions. We fitted the results to the experimental data using the dimensions of the boundary conditions as the fitting parameters. We call $D_{IC}$ the ionic conducting diameter of the pore, $L_{IT}$ the membrane insulating thickness and $d_{DNA}$ the diameter of the insulating DNA molecule at the center of the pore. Actually, rather than fit the value of each pore diameter we set the effective diameter of each nanopore $D_{IC} = D_{pore} - \delta D_{IC}$, to be reduced from each TEM measured pore diameter $D_{pore}$ by $\delta D_{IC}$. The latter is a single fitting parameter used for all pores. This correction allows for the fact that solution ions do not approach the hydrophobic TEM determined pore perimeter due to various chemical size effects. The best fit to the data in Fig. 3a is given by the solid line for which $\delta D_{IC} = (0.65 \pm 0.05)$ nm, $d_{DNA} = (1.93 \pm 0.02)$ nm, and $L_{IT} = (0.37 \pm 0.08)$ nm. The value of $L_{IT}$ determined here is in the range of our previously reported value (0.6 [+0.9, –0.6] nm)[1] and agrees with theoretical predictions[21] for a graphene-water distance of 0.31-0.34 nm. It seems reasonable that the effective diameter reduction $\delta D_{IC}$ be similar to the membrane insulating thickness $L_{IT}$ with any deviations attributed to the energy cost of an ion shedding its hydration shell within the nanopore[22,23], or the consequences of chemical groups bound to the graphene nanopore's edge[22].

The modeled dependence of blockade current vs. pore diameter is shown as the solid curve in Fig. 3a based on the parameters given above. The fit to the data (solid circles) is seen to be rather good and we conclude that the calculation captures the essential physics with the parameters $L_{IT}$ and $D_{IC}$ absorbing all the molecular interaction effects[24].

We define the nanopore's sensitivity as the change in current blockade caused by a change in the diameter of the translocating polymer

$$S(D_{IC}, d_{poly}) = \frac{\partial I_B}{\partial d_{poly}} \approx -\frac{\partial I_B}{\partial D_{IC}}$$

This is the slope of the solid curve as a function of nanopore diameter. The approximation assumes $d_{poly} \approx D_{IC} > L_{IT}$.

The value of $S$ from Fig. 3a, remarkably, exceeds *0.5 nA/Å*, predicting that a change of a tenth Angstrom molecule diameter in our close fitting pore results in a current blockade change of 50 pA in $I_B$ for dsDNA. Similar results should be obtained for ssDNA ($d \approx 1.4$ nm) with a similarly tight fitting nanopore.

Figure 3b shows the current density in and near a 2.5 nm DNA threaded pore using the relevant fitting parameters. The high sensitivity is due to the very concentrated current density in the narrow solution space between the DNA and nanopore periphery. Figure 3c shows a plot of the Sensitivity Gradient along the DNA molecule surface defined by

$$\frac{dS}{dz} = \lim_{\delta D \to 0} \frac{1}{\delta D R} \left( \frac{dV_{D+\delta D}}{dz} - \frac{dV_D}{dz} \right)$$

where $D$ is the polymer diameter, $R$ the pore resistance, $z$ the distance from the symmetry plane of the pore along the surface of the polymer, and $V_D(z)$ is the electrostatic potential along the surface of a polymer of diameter $D$ as a function of distance from the pore $z$. The Sensitivity Gradient identifies where the contributions to S come from along the molecule's length and thus reveals the spatial resolution of the molecule pore system. The Sensitivity Gradient "full width half max" of 0.5 nm in Figure 3c, for the 2.5 nm pore, is the order of expected base separations in ssDNA. The excellent spatial resolution along the molecule is closely connected with the rapid drop in current density on leaving the vicinity of the pore as seen in Fig. 3b.

The sensitivity to molecule size demonstrated here and the high spatial resolution along the molecule's length shown previously[1] are the direct consequence of the atomic thinness of a molecule-hugging graphene nanopore. This, together with the recent proof-of-principle demonstration that very small diameter graphene nanopores can be fabricated with atomic-scale precision[2], suggest that graphene nanopores will be an ideal nanopore sensor.

Achieving ssDNA strand sequencing with graphene nanopores will also require solutions to several of the same problems that have been addressed when utilizing protein pores: the need for ratcheting control of DNA translocation to suppress Brownian molecular motions[25], and the suppression of large 1/f electronic noise that can severely degrade the ionic current signal-to-noise. It remains to be seen if some of the solutions for translocation control used with biological pores[3] may be applied to graphene pores or otherwise engineered[26]. Likewise, further research will be required to determine how to optimally suppress the large observed 1/f electronic noise frequently observed in graphene (Fig. 1e). This noise can be reduced by applying various coatings[16,17], but these coatings increase the length of the nanopore and are incompatible with the high resolution detector strategy reported here. We have achieved noise reduction by simply reducing the free standing area of the graphene membrane (Fig. 1e).

In conclusion we believe the longitudinal sub-nanometer resolution possible in graphene nanopores[1] together with the radial super-sensitivity discovered here bode well for the application of graphene nanopores to many molecular sensing, characterization, and sequencing problems.

**Methods**

The graphene was grown via chemical vapor deposition (CVD) on a copper foil[27], and subsequently transferred[28] over the aperture on a free-standing SiNx film covering a silicon chip, and carefully processed to remove any residual hydrocarbon impurities from the polymer support film used during the transfer process[1]. Individual graphene nanopores were drilled with a JEOL 2010F TEM operating at 200kV. Micro-Raman measurements (model WITec CRM 200) and TEM imaging with atomic resolution (aberration-corrected Zeiss Libra 200, operating at 80keV electron energies) demonstrated that the transferred graphene film was single-layer and largely free of hydrocarbon contaminants. The graphene chip was sealed in a fluidic cell so as to separate two chambers that were subsequently filled with 3 M potassium-chloride solutions, pH 10 or pH12.5. The graphene nanopore was the only path through which ions and DNA molecules (10kBase fragments) could pass between the two chambers. Applying a constant 160 mV bias voltage with Ag/AgCl electrodes in each chamber, we measured the ionic current and DNA translocation through the nanopore using standard electrophysiological methods.

The numerical simulations were performed using the COMSOL Multiphysics finite element solver in 3D geometry, cylindrically symmetric along the axis of the nanopore. A DNA molecule was modeled as a long stiff insulating rod threading the nanopore along its axis. We solved the full set of Poisson–Nerst–Planck (PNP) equations, with the boundary conditions at the graphene corresponding to idealized, uncharged membrane impermeable to ions. In our experimental regime with high KCl concentration and small applied voltage, the PNP solution was found to differ only by few percents from the numerical solution of the Laplace equation with the appropriate electrolyte conductivity. The total ionic current was calculated by integrating current density across the diameter of the nanopore.

**Acknowledgements**

This work was funded by a grant (number R01HG003703) to J.A. Golovchenko and D. Branton from the National Human Genome Research Institute, National Institutes of Health. S. Liu acknowledges support from the State Scholarship Fund of China.

**Author Contributions**

The experiments were designed by S.G., D.B, and J.A.G. Measurements and sample preparation were done by S.G. and S.L. All other activities, including data interpretation, conclusions, and manuscript writing, were carried out collaboratively by S.G, D.B. and J.A.G.

**Additional Information**

The authors declare no competing financial interests. Correspondence and requests for materials should be addressed to J.A.G. (golovchenko@physics.harvard.edu).

**Figure Legends**

**Figure 1**. **(a)** The experimental device with dsDNA translocating through a graphene nanopore. **(b)** Typical current blockades as dsDNA translocates through a 5 nm $SiN_x$ pore and through graphene pores of various diameters (D). **(c)** dsDNA blocked currents through nanopores of different diameters, defined by the open pore current. **(d)** Open pore currents through a series of different diameter pores before addition of DNA. **(e)** Noise power spectral density of a 5 nm graphene nanopore suspended across a 200 x 200 nm $SiN_x$ aperture (graphene 1) and a similar

nanopore across a 20 nm diameter SiN$_x$ aperture (graphene 2) with greatly reduced noise. *A* 5 nm nanopore in SiN$_x$ with no graphene is also shown.

**Figure 2**. **(a & b)** Event plots for different diameter nanopores. Each circle corresponds to a DNA translocation event. Blue, extended, unfolded molecule translocations (pH 10), green, folded translocations (pH 10), and black, single stranded translocations (pH 12.5). Dashed lines represent contours of constant *ecd*. **(c & d)** Single typical event current traces corresponding to one of the events from the event plots in (a) and (b), for folded DNA, unfolded DNA and for single stranded DNA. The event plots and current traces in panels (b) and (d) were all performed with the same nanopore before the DNA was denatured (blue) and after the DNA was denatured (black) by raising the solution pH to 12.5.

**Figure 3.** **(a)** Average current blockade for dsDNA translocations vs. nanopore diameter (solid circles: graphene nanopores, open square: 30nm thick SiN$_X$ pores). Solid and dashed lines are fit to the data using the numerical model. *Inset:* Nanopore conductivity vs. pore diameter. Red line is simulated using best-fit values. **(b)** Calculated current density in a 2.5 nm graphene nanopore traversed by dsDNA, the latter represented by a 1.93 nm cylinder. **(c)** Calculated Sensitivity Gradient along a DNA molecule for the conditions in (b.

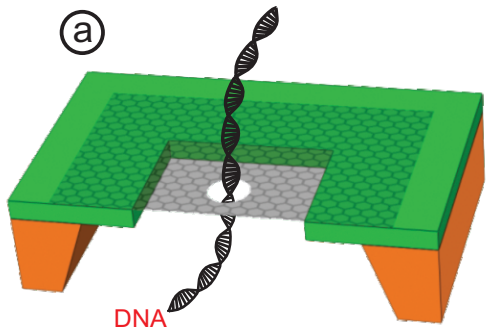
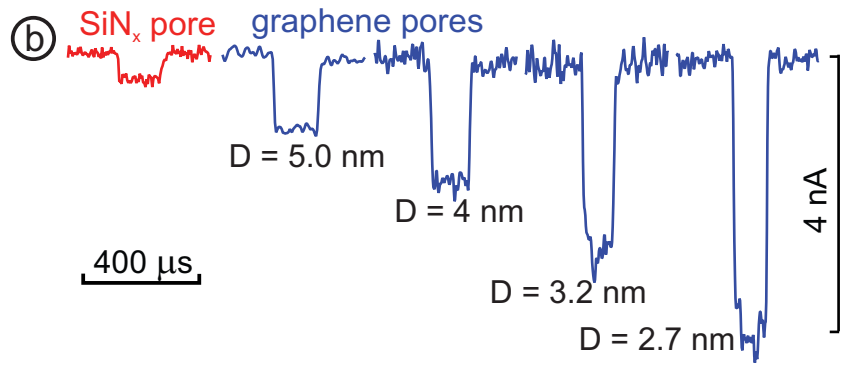
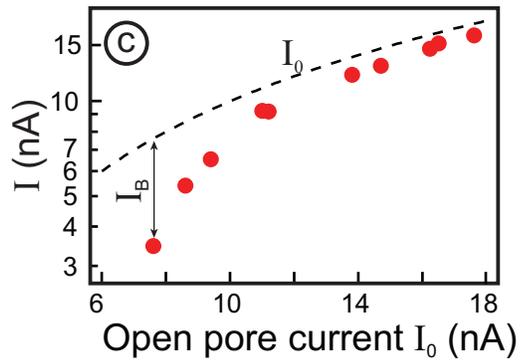
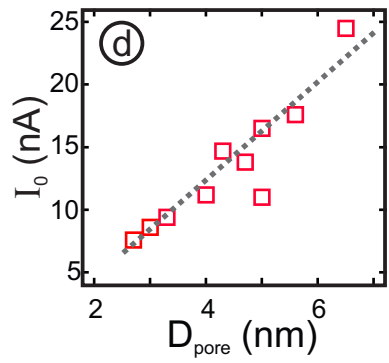
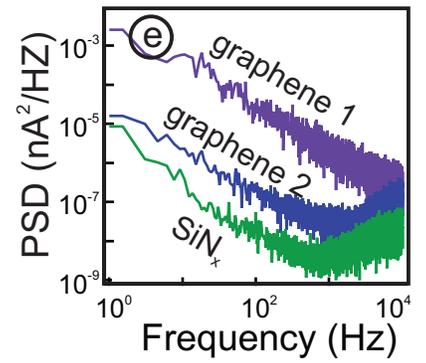

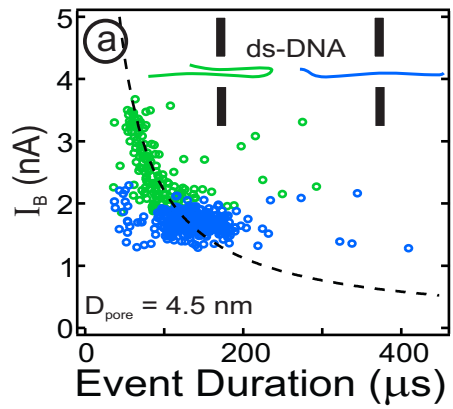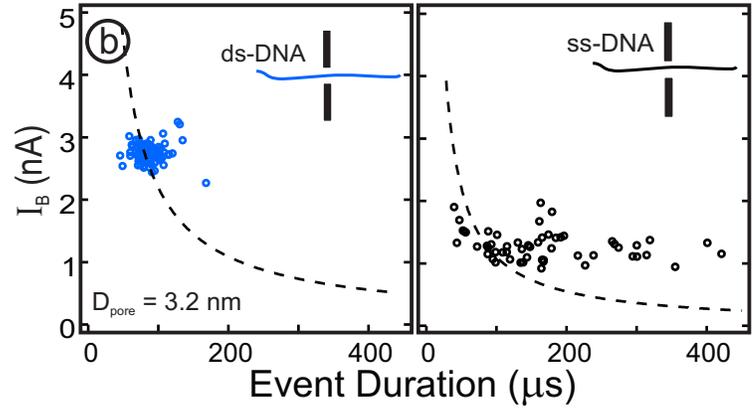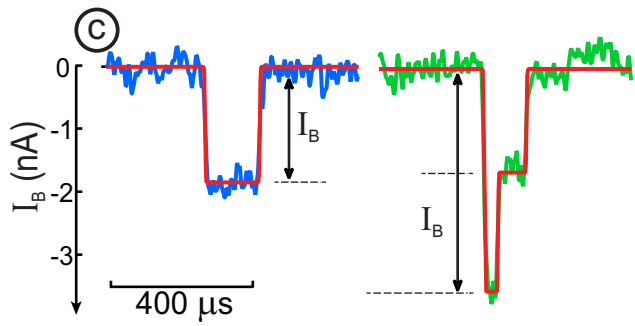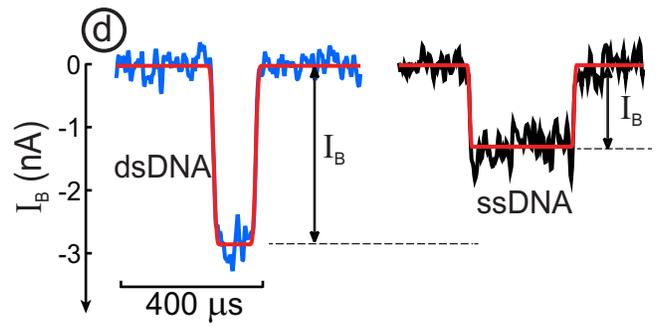

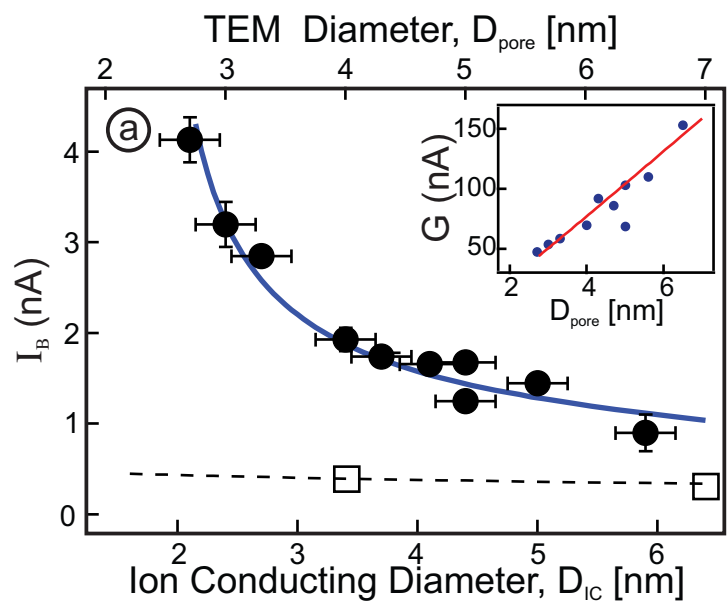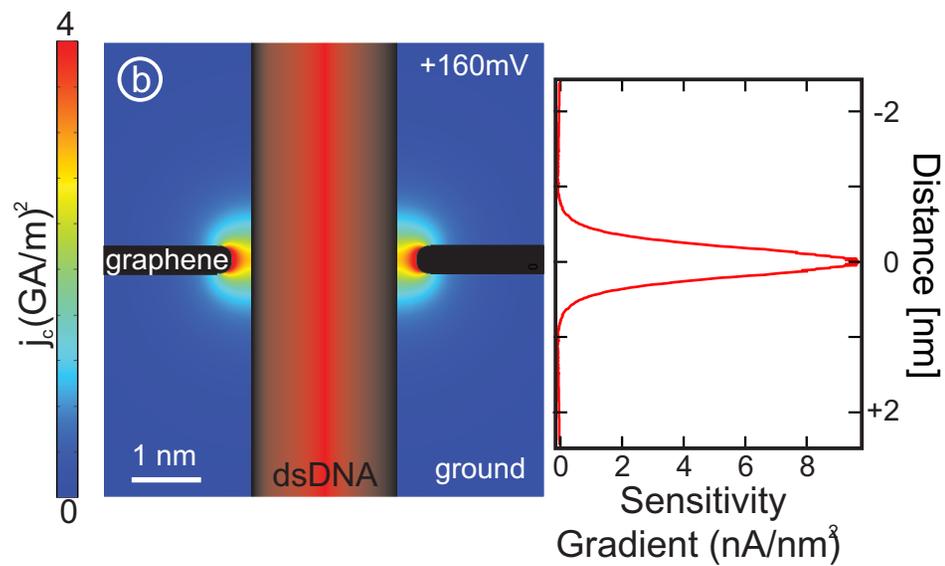